\documentclass[sort&compress,twocolumn,3p]{elsarticle}
\usepackage{mathtools}
\usepackage{gensymb}
\usepackage{graphicx}
\usepackage{dcolumn}
\usepackage{bm}
\journal{Acta Materialia}

\begin{document}
\begin{frontmatter}

\title{Direct Measure of Giant Magnetocaloric Entropy Contributions in Ni-Mn-In}

\author[1]{Jing-Han~Chen\corref{cor1}}
\ead{jhchen@tamu.edu}
\author[2]{Nickolaus~M.~Bruno}
\author[2,3]{Ibrahim~Karaman}
\author[4]{Yujin~Huang}
\author[4]{Jianguo~Li}
\author[1,3]{Joseph~H.~Ross,~Jr.}
\cortext[cor1]{Corresponding author. Tel: +1-979-845-7823}
\address[1]{Department of Physics and Astronomy, Texas A\&M University, College Station, Texas 77843, USA}
\address[2]{Department of Mechanical Engineering, Texas A\&M University, College Station, Texas 77843, USA}
\address[3]{Department of Materials Science and Engineering, Texas A\&M University, College Station, Texas 77843, USA}
\address[4]{School of Materials Science and Engineering, Shanghai Jiaotong University, Shanghai, 200240, China}

\begin{abstract}
Off-stoichiometric alloys based on Ni$_2$MnIn have drawn attention due to the coupled first order magnetic and structural transformations, 
		and the large magnetocaloric entropy associated with the transformations.
Here we describe calorimetric and magnetic studies of four compositions.
The results provide a direct measure of entropy change contributions at low temperatures as well as at the first-order phase transitions.
Thereby we determine the maximum possible field-induced entropy change corresponding to the giant magnetocaloric effect. 
		We find a large excess entropy change above that of the magnetic moments, but only in compositions with no ferromagnetic order in the high-temperature austenite phase.
Furthermore, a molecular field model corresponding to magnetic order in the low-temperature phases is in good agreement,
		giving an entropy contribution nearly independent of composition, despite significant differences in overall magnetic response of these materials.
\end{abstract}

\begin{keyword}
Magnetocaloric effect \sep Heusler alloys \sep Entropy \sep Calorimetry
\end{keyword}
\date{\today}


\end{frontmatter}

\section{Introduction}
The magnetocaloric effect (MCE) is an intrinsic thermodynamic property of magnetic solids, 
	manifested as an adiabatic temperature change or an isothermal entropy change due to application of a magnetic field.
Materials showing a large MCE have been a source of growing interest because of their potential for environmentally friendly and energy efficient replacement of vapor-compression refrigeration \cite{0034-4885-68-6-R04,GschneidnerJr2008945}.
Giant MCE is based on a first-order phase transition and has been observed in materials including the Ni-Mn based Heusler alloys \cite{:/content/aip/journal/apl/85/19/10.1063/1.1808879,krenke2005inverse,:/content/aip/journal/apl/88/12/10.1063/1.2187414,:/content/aip/journal/jap/101/5/10.1063/1.2710779,Liu2012514,Bennett201234,PhysRevB.75.104414,:/content/aip/journal/apl/89/18/10.1063/1.2385147} discussed here,
	  as well as Gd$_5$Si$_{2-x}$Ge$_x$ materials \cite{PhysRevLett.78.4494},
	  MnAs-based compounds \cite{Wada2003114,tocado2006adiabatic,tegus2002transition} 
	  and LaFe$_{13-x}$Si$_x$ related compounds \cite{:/content/aip/journal/apl/78/23/10.1063/1.1375836,:/content/aip/journal/apl/101/16/10.1063/1.4760262}.
Ni-Mn-In alloys exhibit a large response, and have magnetic and structural properties that depend very sensitively on the composition and preparation conditions \cite{0953-8984-21-23-233201}.
In this report, we examine four different Ni-Mn-In compositions to better understand the relationship of magnetic and structural properties vs. composition.

\begin{figure}
\includegraphics[width=0.6\columnwidth]{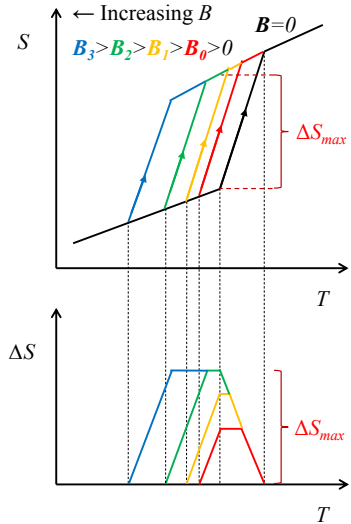}
\caption{Schematic representation of the total entropy in different applied fields along with the field-induced isothermal entropy change (lower plot) for inverse MCE materials in the vicinity of a first-order phase transition.}
\label{intro}
\end{figure}

One of the common physical quantities used to characterize MCE materials is the magnetic field-induced isothermal entropy change.
This quantity can be explored both by calorimetry and indirectly through magnetization \cite{PhysRevB.77.214439,0953-8984-21-7-075403,Suzuki2010693,0022-3727-45-25-255001,Bruno201466,:/content/aip/journal/jap/116/20/10.1063/1.4902527,tocado2009entropy}.
As Fig.~\ref{intro} illustrates, the entropy change intrinsic to the first-order transformation is the maximum possible field-induced entropy change ($\Delta S_{max}$), which can be explored directly through calorimetry in zero field even though this quantity may not be necessarily accessible through measurements of $\Delta S$ in available fields.
Aside from the practical importance there is considerable interest in understanding the underlying physical processes that contribute to this behavior \cite{GschneidnerJr2012572,:/content/aip/journal/apl/95/17/10.1063/1.3257381,entel2013complex}.
 In this report, we examine the various contributions to the total entropy in several Ni-Mn-In alloys, including an examination of the contributions to $\Delta S_{max}$. 

We describe results based on magnetic and calorimetric measurements.
 These results provide a consistent picture of the magnetic behavior of the low temperature (martensite) phases,
 dominated by magnetic order based on interacting local moments.
Furthermore, the entropy jump at the transformation, equivalent to $\Delta S_{max}$, is shown to be very sensitive to the composition.
Relative to the magnetic contribution, 
		 a significant excess is found in compositions exhibiting magnetic order in the austenite (high-temperature) phase,
		 implying an apparent magnetoelastic coupling tied to the magnetic order of the austenite.

\section{Experiment}
\subsection{Sample Preparation}
Bulk polycrystalline Ni-Mn-In alloys were prepared using arc melting in a protective argon atmosphere from 99.9\% pure constituents.
This includes the sample used for measurements reported previously \cite{:/content/aip/journal/jap/116/20/10.1063/1.4902527} along with 3 other compositions prepared by identical methods.
The samples were homogenized at 1173 K for 24 hours under argon atmosphere and then quenched to room temperature in water.
These heat treating conditions were found to result in sharp first-order transitions and small thermal hysteresis.
Since the alloy was quenched above the order disorder temperature \cite{Recarte20121937}, $B2$ crystallographic symmetry will dominate in these samples with Ni on one sublattice and Mn and In randomly occupying the other, rather than the Heusler-type $L2_1$ ordering obtained at lower temperatures.

Electron microprobe measurements were carried out using wavelength dispersive spectroscopy methods on a Cameca SX50, equipped with four wavelength-dispersive x-ray spectrometers.
Among the four Ni-Mn-In compositions, the final compositions were found to be Ni$_{49.54}$Mn$_{36.12}$In$_{14.34}$ (the sample used in the previous report \cite{:/content/aip/journal/jap/116/20/10.1063/1.4902527}), Ni$_{49.9}$Mn$_{35.7}$In$_{14.4}$, Ni$_{49.53}$Mn$_{35.22}$In$_{15.22}$ and Ni$_{47.22}$Mn$_{38.45}$In$_{14.33}$.
We label these samples as A, B, C and D, respectively, as shown in Table~\ref{comptable}.
\begin{table*}
\centering
\begin{tabular}{l|cccccc}
\hline\hline
label&WDS composition&$J$&$T_c$(K)&$T_{mh}$(K)&$\gamma$(J/mole K$^2$)&$\theta_D$(K)\\
\hline
A	&Ni$_{49.54}$Mn$_{36.12}$In$_{14.34}$	&2.00	&292	&347     	&0.0124     	&315\\
B	&Ni$_{49.88}$Mn$_{35.70}$In$_{14.42}$	&1.99	&310	&333     	&0.0128     	&318\\
C	&Ni$_{49.53}$Mn$_{35.22}$In$_{15.22}$	&2.00	&323	&299		&0.0117     	&316\\
D	&Ni$_{47.22}$Mn$_{38.45}$In$_{14.33}$	&2.00	&298	&257     	&0.0163     	&316\\
\hline\hline
\end{tabular}
\caption{Ni-Mn-In compositions (in at. \%) and the corresponding experimental results.
		$J$ and $T_c$ are from the high-temperature Curie-Weiss fits, and 
		$T_{mh}$ (martensitic transition temperature) is the maximum position of the specific heat measured while heating.}
\label{comptable}
\end{table*}

\subsection{Measurement Methods}
Iso-field magnetic measurements were carried out using a Quantum Design Magnetic Property Measurement System.
The temperature-dependent results in 0.05 T shown in Fig.~\ref{mt} include prominent features due to the first-order martensitic phase transitions falling between 200 and 350 K.
As shown below, the smaller responses in the A and B compositions come about since the austenite (high-temperature) phase for these cases are paramagnetic rather than ferromagnetic.

Besides the magnetic measurements, calorimetric measurements were performed using a Physical Property Measurement System manufactured by Quantum Design.
Samples were affixed using a thin layer of grease to the small sample platform of a standard puck, with the platform connected by thin wires to the body of the puck.
For temperatures away from the structural transition, the so-called 2-$\tau$ method \cite{:/content/aip/journal/rsi/68/1/10.1063/1.1147722} was used.
Since this method is not valid for a first order transition, we used the modified method described previously \cite{:/content/aip/journal/jap/116/20/10.1063/1.4902527}, by which we obtain results consistent with the 2-$\tau$ method outside the transition temperature region.

\section{Magnetization Analysis}
\begin{figure}
\includegraphics[width=\columnwidth]{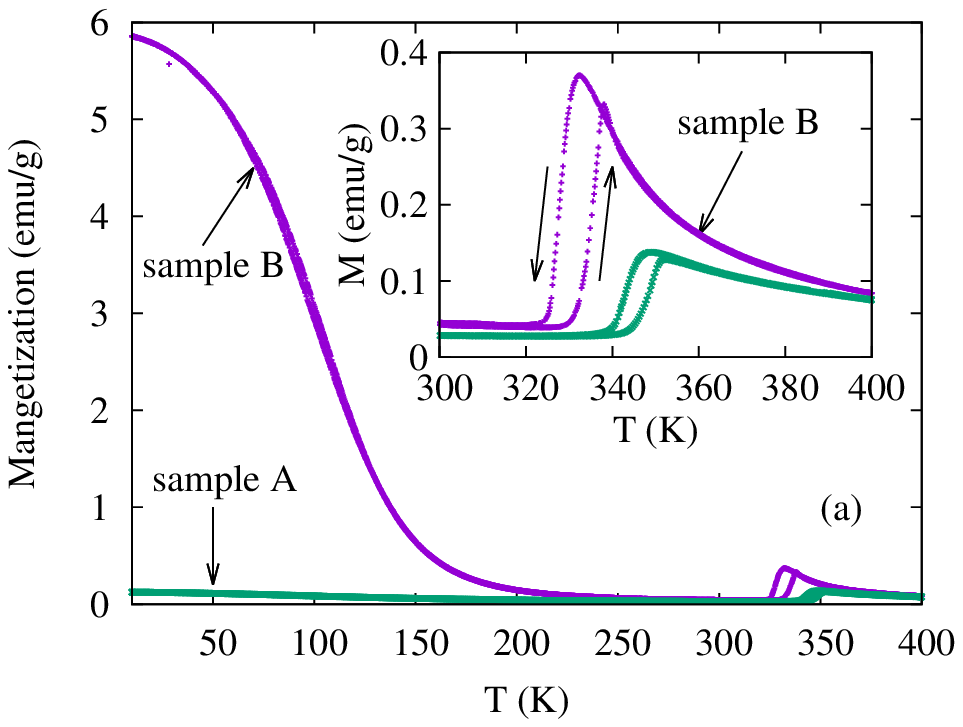}
\includegraphics[width=\columnwidth]{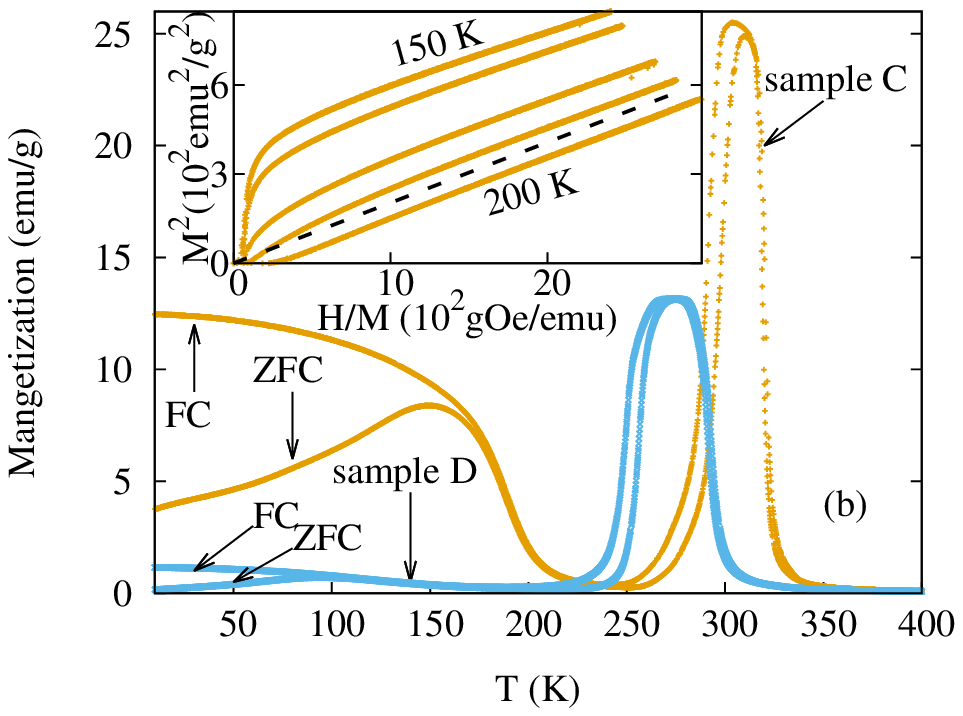}
\caption{Temperature dependence of the magnetization for the Ni-Mn-In samples at 0.05 T, including (a) A and B samples and (b) C and D samples.
 All data include results for both heating and cooling processes, as shown by arrows for sample B in the inset of the upper figure.
 For samples C and D, the low-temperature bifurcation corresponds to field-cooled (FC) and zero-field cooled (ZFC) measurements as shown.
 An Arrott plot for sample C is shown in the lower inset. These curves were measured at temperatures 150, 160, 180, 190, and 200 K.
 The dashed line extending through the origin corresponds to 194K, the approximate position of the second order transformation as described in the text.}
\label{mt}
\end{figure}

The magnetization results shown in Fig.~\ref{mt} for the four samples show sharply-defined martensitic transitions from a paramagnetic/ferromagnetic austenite to an antiferromagnetic (or low-moment) martensite upon cooling, with the reverse transition observed on heating. 
These transitions are in good agreement with reported phase diagrams for similar compositions \cite{0953-8984-21-23-233201}, with the A and B compositions remaining paramagnetic at all temperatures above the martensitic transition, while C and D exhibit a ferromagnetic Curie temperature ($T_c$) within the austenite phase.
As shown below, the magnetization can be fitted very well in terms of fixed local moments residing on the Mn ions.

The Curie-Weiss law,
\begin{equation}
M=n\frac{N_A}{3k_B}\mu_{eff}^2\frac{H}{T-T_c},
\label{weiss}
\end{equation}
where $\mu_{eff}=g\mu_B\sqrt{J(J+1)}$, $T_c$ is the Curie temperature and $g=2$,
was used to fit the high temperature magnetization curves in the paramagnetic region i.e. above $T_c$ for the C and D compositions. 
This fitting assumed that the density of magnetic moments ($n$) is identical to the manganese ion density.
For the four compositions, the results yield $J$ very close to 2 as displayed in Table~\ref{comptable}.
The consistency of these results with a local magnetic moment of $gJ\mu_B=4\mu_B$ per Mn ion also agrees with computed results \cite{Li201335,0953-8984-21-23-233201,PhysRevB.86.214205,:/content/aip/journal/apl/97/24/10.1063/1.3525168}.
It appears that in all cases the magnetism corresponds to local moments which are indirectly coupled through RKKY interactions \cite{kittel1987quantum}.
Similar behavior has previously been identified in other X$_2$MnY Heusler systems \cite{0305-4608-6-8-007,PhysRevB.28.1745}, and we show below that this result helps in analyzing the specific heat results.

Fitted $T_c$ values are also given in Table~\ref{comptable}.
Notice that for the A and B samples, these values are lower than the martensite transition temperatures corresponding to an austenite phase which remains paramagnetic.
The C and D compositions become ferromagnetic upon heating before becoming paramagnetic.

\begin{figure}
\includegraphics[width=\columnwidth]{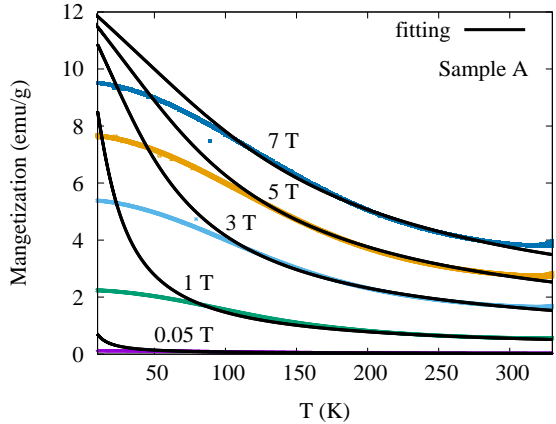}
\caption{Temperature dependence of the martensite phase magnetization for Ni$_{50}$Mn$_{36}$In$_{14}$ (sample A) at magnetic fields 0.05, 1, 3, 5 and 7 T.
The solid curves are from fitting a single set of parameters for a combination of paramagnetic clusters and an AF matrix.}
\label{superpara}
\end{figure}

At low temperatures, the magnetization of each sample tends to saturate at a value lower than $4\mu_B$ per Mn showing the antiferromagnetic or ferrimagnetic behavior of the martensite phases.
For the case of sample A, good agreement with the $M$ vs. $T$ curves above about 100~K (Fig.~\ref{superpara}) was obtained by assuming that a small fraction of the magnetic moments form small superparamagnetic clusters, and the remainder forms an antiferromagnetic (AF) matrix at all temperatures.
Such a model has been used previously for Ni-Co-Mn-Sn Heusler alloys \cite{Cong20125335}.
Since the N\'{e}el temperatures of our materials are high (about 500~K, as shown below), we treated the antiferromagnetic susceptibility as nearly temperature-independent at these temperatures.
Therefore, the martensite magnetization curves were fitted by a sum of paramagnetic and antiferromagnetic contributions
\begin{equation}
M=NgJ^\prime\mu_BB_{J^\prime}\left(\frac{gJ^\prime\mu_BH}{k_BT}\right)+\chi_{AF}H,
\end{equation}
where $B_{J^\prime}$ is a Brillouin function for the clusters each with total spin $J^\prime$ and $\chi_{AF}$ is the constant antiferromagnetic susceptibility.
In the fit (from 100 K to 300 K, with the single fit including data for all fields), we obtained $J^\prime= 26$, and a cluster density $N =3.9\times 10^{-5}$ mole/g giving the results shown in the figure.
Assuming the clusters to be locally ferromagnetic with fully-aligned Mn ions, each with $J = 2$ consistent with the Curie-law fitting,
		 this corresponds to 9.3\% of the Mn atoms comprising the superparamagnetic clusters and the remaining Mn atoms contained in the AF matrix.

This fit also provides information about the antiferromagnetic state of the bulk, since, as is well established in that case, similar to Eq.~\ref{weiss}, $\chi_{AF}=N_An\mu_{eff}^2/(6k_BT_N)$ at the N\'{e}el transition temperature $T = T_N$,
within a molecular field model, while for a powder sample the susceptibility will drop slowly to 2/3 of this value approaching $T = 0$.
Equating the fitted $\chi_{AF}=9.37\times 10^{-6}$emu/g G to the value at $T_N$, and assuming the AF matrix includes 91\% of the Mn ions as established in the fit, we obtain $T_N=610$~K.
Renormalizing this value to the small decrease in $\chi_{AF}$ in the fitted range gives $T_N\approx 500$~K.
As shown below, this extrapolated $T_N$ is also consistent with the measured entropy of the AF phase for this composition.

At lower temperatures in Fig.~\ref{superpara} the departure of magnetization from the fitted curves is likely due to super-spin glass behavior similar to what has been established in related alloys \cite{Cong20125335}, combined with blocking behavior of individual clusters.
Indeed, the sample A curve of Fig.~\ref{mt}(a) has a FC vs. ZFC bifurcation near 100 K, not easily distinguished due to the vertical scale, as would be expected for such a situation.
Corresponding behavior is seen more clearly for sample D in Fig.~\ref{mt}(b).

The analysis establishing superparamagnetic behavior in an AF matrix works particularly well for sample A since there is a large martensite temperature region over which this behavior could be fitted.
Sample D behaves in a similar way, and we also found that the low-temperature $M$-$T$ curves vs. field for this sample (not shown) saturate to approximately 10\% of the full Mn moment.
This indicates the presence of a comparable density of superparamagnetic clusters, although the smaller temperature window makes a quantitative estimate of the contribution of the antiferromagnetic matrix to the susceptibility of this sample more difficult.
For sample B the response is qualitatively similar, but with a larger magnetization developing at low temperatures [Figure~\ref{mt}(a)].
We have not fully characterized the magnetic response for this sample, although from the specific heat results (below) we deduce that the degree of magnetic order in this sample is similar to that of the others, so its low-magnetization behavior above about 200 K indicates that the martensite phase is also essentially antiferromagnetic.
By contrast, sample C exhibits a clear signature of an additional magnetic phase transition at about 194 K;
this is shown by the series of Arrott plots in the inset of Figure~\ref{mt}(b), which demonstrate typical results of second order transformation to a phase exhibiting spontaneous magnetization at low temperatures.

The Arrott plots for sample C follow the standard type corresponding to mean-field critical exponents \cite{PhysRevLett.19.786,PhysRevB.34.3456}.
 Note that for a composition range close to that of this sample, a low-temperature transformation has indeed been previously reported within the martensite phase \cite{0953-8984-21-23-233201},
 although this transformation has often been referred to as a paramagnetic to ferromagnetic transformation.
 We find that the 7 T saturation moment for this sample is 45 emu/g, or about 30\% of what is expected from complete alignment of the Mn local moments.
 In addition, the calorimetric results described below indicate that this is an order-to-order transformation of the magnetic moments,
 so tentatively we conclude that it is a canting-type transformation within the antiferromagnetic martensite phase of this sample.
 With decreasing In content this transformation rapidly falls away to zero temperature \cite{0953-8984-21-23-233201}, explaining why the other samples do not show the additional low-temperature magnetic phase.

\section{Specific Heat}
\begin{figure}
\includegraphics[width=\columnwidth]{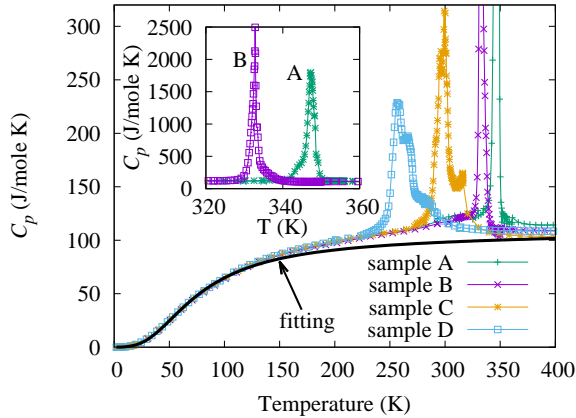}
\caption{Temperature dependent heating curve specific heat results for the four Ni-Mn-In samples. 
		Fitting curve is phonon + electron result for sample A; similar curves obtained for the other samples.
		Inset: full-scale results for the A and B compositions.}
\label{Cp}
\end{figure}

In order to gain a systematic understanding of the entropy, we performed calorimetric measurements from 1.8 K to 400 K.
Figure~\ref{Cp} shows heating curve results (martensite$\to$austenite) for the four Ni-Mn-In samples.
The temperatures where the heating curve specific heat reaches a maximum are denoted as $T_{mh}$ and listed in Table~\ref{comptable} as representing the martensitic transition temperature.
These transitions and the Curie temperatures are consistent with the magnetic results (Fig.~\ref{mt}).
The cooling curves also give similar results, however the analysis below concentrates exclusively on the heating curves.

Owing to the diffusionless character of the martensitic transition, configurational contributions are believed to be largely absent \cite{:/content/aip/journal/jap/93/10/10.1063/1.1556218} so that the specific heat can be simplified as a combination of electronic, vibrational and magnetic parts:
\begin{equation}
C_{total}\approx C_{el}+C_{vib}+C_{mag}.
\label{ctot}
\end{equation}
The electronic contribution can be described by the known linear-$T$ behavior \cite{mizutani2001introduction},
\begin{equation}
C_{el}(T)\cong\left(\frac{\pi^2}{3}\right)k_B^2D(E_F)T=\gamma T,
\end{equation}
where $D(E_F)$ is the density of states at the Fermi energy.
The vibrational contribution \cite{mizutani2001introduction} based on a Debye model is
\begin{equation}
C_{vib}(T)\cong 9R\left(\frac{\theta_D}{T}\right)^3\int_0^{\theta_D/T}\frac{x^4e^x}{(e^x-1)^2}dx
\label{debye}
\end{equation}
where $R$ is the ideal gas constant and $\theta_D$ is the Debye temperature.
We fitted these two terms below 100~K, and the extrapolated curves with $\theta_D$ and $\gamma$ kept constant were used to analyze the high-temperature behavior as discussed below.
The fits yielded $\theta_D$ ranging from 315~K to 318~K and $\gamma$ ranging from 0.0117 to 0.0163 J/mole K$^2$, shown in Table~\ref{comptable}.

\section{Entropy Analysis}
In thermodynamics, total entropies have an abrupt change at the temperature where a first-order phase transition happens, as shown in Fig.~\ref{intro}.
With the electronic and vibrational contributions obtained as described above, using Eq.~\ref{ctot} we can obtain the excess contribution to the entropy (per mole Mn) as
\begin{equation}
\begin{split}
S_{excess}&=\frac{1}{n}\int\frac{C_{total}-C_{vib}-C_{el}}{T}dT\\
&=S_{mag}+S^\prime\leq R\ln(2J+1)+S^\prime,
\end{split}
\label{smag}
\end{equation}
where $S_{mag}$ is the entropy due to magnetism and $S^\prime$ represents possible additional nonmagnetic entropy.
The last inequality expresses the well-known limit for the magnetic entropy \cite{mayer1940statistical} in which as shown above Mn ions can be treated as local moments with $J=2$.

\begin{figure}
\includegraphics[width=\columnwidth]{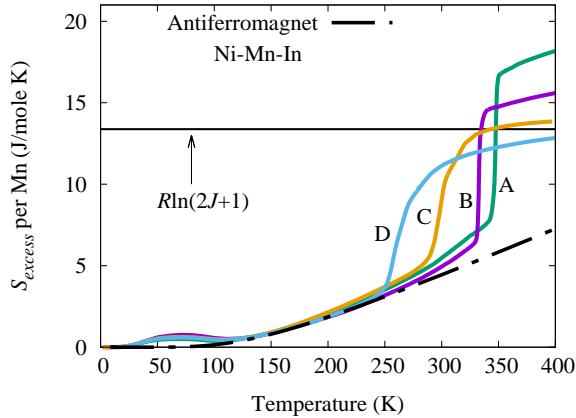}
\caption{Excess entropy of samples A-D per mole Mn atom, obtained from Eq.~\ref{smag}. As can be seen, samples A and B exceed the classical magnetic entropy limit, shown as the horizontal line.
Dashed curve represents AF contribution with $T_N=540$ K.}
\label{mgentropy}
\end{figure}

The results are represented in Fig.~\ref{mgentropy}, where the horizontal line represents the magnetic entropy limit.
The small entropy bump before 100 K is believed to be due to the limitation of the constant Debye temperature approximation in which the phonon density of modes at low frequencies are underestimated \cite{gopal1966specific}.

For C and D compositions with $T_{mh}/T_c<1$,	the samples go to a ferromagnetic high-\textit{T} austenite phase as remarked above.
The entropy contribution for these samples appears to be quite close to the classical magnetic limit as shown in Fig.~\ref{mgentropy}.
On the other hand, for samples A and B where $T_{mh}/T_c>1$, the martensite phases transform directly to a paramagnetic austenite.
In these cases, the remaining entropy goes significantly beyond the magnetic entropy limit.
This is the first time in the literature as far as we are aware of for observation of the excess entropy directly from calorimetric measurements.

Below $T_{mh}$ the gradual magnetic entropy changes in the martensite phase can be understood as due to order-disorder within an antiferromagnetic state \cite{0034-4885-17-1-307}.
This contribution for different $J$ and $T_N$ can be estimated numerically from molecular field theory.
We showed above that an antiferromagnetic case applies very well to the sample A magnetization, however the mean-field analysis we have used applies in general to any ordered magnetic state, since the entropy will be zero in the fully ordered state whether ferrimagnetic or antiferromagnetic and the increase with temperature follows a standard curve \cite{0034-4885-17-1-307} as the on-site magnetization becomes reduced.
The computed result from molecular field theory gives good agreement with our experiment, as shown by the dashed line in Fig.~\ref{mgentropy}.
This curve represents $J=2$ antiferromagnetism with N\'{e}el temperature set to 540~K, and gives qualitatively good agreement with the magnetic entropy in the martensite phase for all four samples.

Note that this includes sample D, which exhibits a clear change in magnetic order at 194 K as noted above.
In some cases it has been speculated that similar compositions go over to a magnetically disordered state due to frustrated magnetic interactions,
   however from the extracted entropy it is clear that all of these samples maintain a similar degree of magnetic order while in the martensite phase.

\section{Discussion}
There has been considerable interest in understanding the local nature of the magnetic moments in these materials,
	 and it is interesting that \textit{ab initio} calculations typically indicate a small nonzero moment on Ni \cite{0953-8984-21-23-233201,Li201335,PhysRevB.86.214205,:/content/aip/journal/apl/97/24/10.1063/1.3525168,PhysRevB.87.144412}.
On the other hand, the magnetic and low-temperature calorimetric results can be understood consistently in terms of a $J = 2$ moment associated with each Mn.
The results for samples C and D with entropy contribution above the transition close to what is expected based on this picture, provides further evidence for this.
 Even very small Ni moments would add a large contribution to $S_{mag}$ if acting independently, similar to the $R\ln(2J+1)$ term in Eq.~\ref{smag}.
 Thus, it appears that the Ni contribution is strongly coupled to the Mn ions and these ions can together be regarded as constituting the overall local moment rather than as thermodynamically independent spins.

The electronic entropy change due to $\gamma$ differences between the martensite and austenite phases is normally believed to be much smaller than the contribution of structural deformations.
 It has been inferred in recent studies of Hesuler alloys \cite{Fraga1991199,:/content/aip/journal/jap/93/10/10.1063/1.1556218,0022-3727-43-5-055004} that the electronic contribution (typically $\gamma$ = 5-10 mJ/mole~K$^2$) is considerably smaller than the vibrational contribution.
Attributing all of the observed excess entropy to an electronic contribution, the change of $\gamma$ must be 14 mJ/mole~K$^2$ in sample A, 
which means a doubling of $\gamma$ from the martensite, while a change of 5 mJ/mole~K$^2$ is required in sample B.
 Also, a recent measurement of Ni-Co-Mn-In in high magnetic fields \cite{PhysRevB.90.214409} indicated that in low temperatures, the low-temperature austenite phase induced by high magnetic fields has an electronic $\gamma$ that differs by a relatively small amount (4 mJ/mole~K$^2$) compared to the low-temperature martensite phase.
 Thus it seems likely that the excess contributions obtained by direct measurement of the total entropy, particularly the large contribution in sample A, are due predominantly to phonons.

The lattice entropy contribution (per mole) can be obtained from the integration of Eq.~\ref{debye},
\begin{equation}
\begin{split}
&S_{vib}(T)=\int\frac{C_{vib}(T)}{T}dT\\
&=-3R\ln\left[1-e^{-\frac{\theta_D}{T}}\right]+12R\left(\frac{T}{\theta_D}\right)^3\int_0^{\theta_D/T}\frac{x^3}{e^x-1}dx.
\end{split}
\label{debyeentropy}
\end{equation}
If we relax the criterion that $\theta_D$ remains constant, and instead assume that $\theta_D$ changes discontinuously during the structural transformation to the austenite phase, the vibrational entropy will also change discontinuously.
In this case the austenite contribution will be represented by Eq.~\ref{debyeentropy}, with the integration including the revised $\theta_D$ extending over the entire range.
This is the most likely scenario for the excess entropy contributions, since the martensitic transition involves a significant unit cell deformation.

If we attribute the entire additional entropy change to vibrational contributions, the corresponding Debye temperatures change from 315 K to 300 K going across transformation in sample A, and from 318 K to 312 K in sample B.
Note that both of these samples, as well as sample C, correspond to compositions for which the martensite phase is believed to have the 10M structure \cite{0953-8984-21-23-233201}, therefore the large excess is not correlated to the structural symmetry itself.
In addition, the martensite phases of all of these compositions feature antiferromagnetic phases with quantitatively quite similar exchange interaction strengths, judging from the similarity in the observed low temperature magnetic entropy, modeled above according to a molecular field theory with the same $T_N$.
The most obvious difference in these materials is the absence of ferromagnetism in the samples A and B austenite phases, which appears to correlate with the large excess entropy jump.
Thus it appears that the relative lattice softening of the paramagnetic austenite phases plays a large role in the magnitude of the magnetocaloric entropy change. 

Returning to size of the entropy jump itself, note that, compared to the magnetic limit $R\ln(2J + 1)$, the entropy increases by a magnitude equal to 70\% of this value in sample A, and 63\% in sample B, within the very narrow temperature range of the transitions.
The relative sizes are apparent in Fig.~\ref{mgentropy}. 
In sample A the jump is also larger than the magnetic entropy is predicted to be available based on the fitted antiferromagnetic fit (dashed curve in Fig.~\ref{mgentropy}).
Thus these values are indeed quite large relative to the available magnetic entropy.
We show that this occurs in a strongly composition-dependent manner, with the paramagnetic phases exhibiting a large enhancement.
This may be connected to recent results based on neutron scattering studies \cite{PhysRevB.92.140406} of a NiMnCoIn composition, in which it was shown that the contribution of phonon modes becomes considerably enhanced when going through the ferromagnetic-paramagnetic transition in the austenite.
To optimize the magnetocaloric response it appears that exploiting alloys having a martensitic transition directly to the paramagnetic phase, but very close to the ferromagnetic part of the phase diagram so that the magnetic coupling is enhanced, can be beneficial.

\section{Conclusion}
Investigation of several compositions of NiMnIn-based Heusler alloys provided a direct measure of thermodynamic properties of the coupled magnetic-structural transformations.
These studies, combining results of magnetic and calorimetric measurements, could be analyzed to give a consistent measure of the magnetic ordering within the low-temperature martensite phase.
For the composition with highest ordering temperature, the results correspond to a predominantly antiferromagnetic martensite phase with N\'{e}el temperature near 500 K.
Study of the excess entropies after accounting for phonon and electronic contributions indicates qualitatively similar magnetic interactions within the other compositions.
The jump in entropy at the first-order structural transformation temperatures was also analyzed and found to be strongly composition-dependent, with compositions no ferromagnetic order in the austenite exhibiting significant excess entropy over the purely magnetic contribution.
The results appear to correlate to the presence of magnetoelastic coupling in the high-temperature austenite phase, and may provide further assistance in design of materials for practical magnetocaloric applications.

\section*{Acknowledgement}
This material is based upon work supported by the National Science Foundation under Grant No. DMR-1108396, and by the Robert A. Welch Foundation (Grant No. A-1526).

\bibliographystyle{elsarticle-num}
\bibliography{elsarticle}

\end{document}